\documentclass[prd,showpacs,showkeys,twocolumn,floatfix]{revtex4} 
\usepackage[]{epsfig}

\usepackage{graphicx}
\usepackage{bm}

\def\lsim{\mathrel{\rlap{\lower4pt\hbox{\hskip1pt$\sim$}}
   \raise1pt\hbox{$<$}}}
\def\gsim{\mathrel{\rlap{\lower4pt\hbox{\hskip1pt$\sim$}}
    \raise1pt\hbox{$>$}}}

\begin{document}


\title{ Final State Interactions in Decays of the Exotic  $\pi_{1}$ Meson } 

\author{ Nikodem J. Poplawski, Adam P. Szczepaniak, J.T.~Londergan }
\affiliation{ Physics Department and Nuclear Theory Center \\
Indiana University, Bloomington, Indiana 47405 }

\date{\today}

\begin{abstract}

We analyze the role of  final state interactions (FSI) in  decay of the 
lighest exotic meson, with $J^{PC}=1^{-+}\,\,\pi_{1}$. We use the 
relativistic Lippmann-Schwinger equation for two coupled $\pi b_{1}$ 
and $\pi\rho$ channels. The first one is the predicted dominant decay mode 
of the $\pi_{1}$, whereas in the other a narrow $\pi_1(1600)$ exotic signal 
has been reported by the E852 collaboration. The FSI potential is 
constructed, based on the $\omega$ meson exchange between the two channels. 
We find that this process introduces corrections to the $\pi_{1}$ widths 
of the order of only a few MeV. Therefore, we conclude that a substantial 
$\pi\rho$ mode cannot be generated through level mixing. 


\end{abstract}


\maketitle

\section{Introduction}

Exotic mesons may provide a unique experimental handle on gluonic excitations. 
By definition, spin exotic mesons have quantum numbers, spin ($J$), parity 
($P$) and charge conjugation ($C$), which cannot be obtained by combining 
quantum numbers of the valence quark and antiquark alone. Thus other  
degrees of freedom, in addition to the valence components, must be present 
in the leading Fock 
space component of an exotic meson to generate exotic combinations of 
$J^{PC}$, {\it e.g.} $J^{PC}=0^{--}, 0^{+-},1^{-+},2^{--},\cdots$. Since multiquark states are expected to fall apart, it is expected that the 
leading components of exotic mesons will contain  excitation  of the gluonic 
field in the presence of just the valence $Q{\bar Q}$ pair. This is confirmed 
by lattice simulations~\cite{Juge:1999ie, Juge:1997nc}
which find that the lightest exotic meson has  
$J^{PC} = 1^{-+}$ and overlaps with the state created from the vacuum by 
application of operators with $Q$, ${\bar Q}$ and gluon field (link) 
operators. The  mass of this lightest exotic is expected between 
$1.6-2\mbox{ GeV}$~\cite{Bernard:2003jd, Manke:1998yg, Bernard:1997ib,Lacock:1996ny} with the uncertainty in the lattice estimates coming 
from chiral extrapolations~\cite{Thomas:2001gu}. There are also estimates of exotic meson 
properties based on the Born-Oppenheimer approximation which indicate that 
the gluon, if treated as a constituent particle in the ground state exotic 
meson, has to carry one unit of orbital angular  momentum with respect 
to the $Q{\bar Q}$~\cite{Swanson:1998kx}.  

Whether exotic mesons have been seen experimentally 
is an open question. Some tantalizing candidates exist in particular  
in the $\eta'\pi$ and $\rho\pi$ decay channels~\cite{Ivanov:2001rv,Adams:1998ff} . The $J^{PC}=1^{-+}$ partial 
wave in the $\eta'\pi$ system measured in $\pi^- p \to \eta' \pi^- p$ is 
as strong as the $J^{PC}=2^{++}$ wave from the $a_2$ resonance decay. 
Unfortunately, the exotic wave peaks at $M_{\eta'\pi} \sim 1.6\mbox{ GeV}$ 
where there are no other well established resonances decaying to 
$\eta'\pi$. Therefore extraction of the phase of the exotic wave, needed to 
establish its dynamical character, depends on model assumptions regarding the 
behavior of other waves. With the GlueX/Hall D experiment one should be able 
to circumvent  this problem by measuring several decay modes with high 
statistics and performing a systematic coupled channel study together with 
the analysis of production mechanisms.  If current experimental results do 
indeed correspond to elementary QCD exotic mesons, then there might be 
a potential discrepancy between model predictions and the experimental data. 
Most models predict that exotic mesons should primarily decay to "exotic" 
channels dominated by two meson final states with one of them being an 
$S$-wave and the other $P$-wave meson~\cite{IKP,Kokoski:1985is,CP,PSS,N} {\it e.g.} $b_1\pi$, $f_1\pi$, 
$K_1,\pi$, while the experimental sightings come for analysis of "normal" 
decay channels {\it i.e.} with two $S$-wave mesons, like the $\eta'\pi$ 
and the $\rho\pi$ modes.  In this paper we examine the effects of final 
state rescattering between mesons as a potential mechanism for shifting 
strength from one channel to another. 

In a previous paper~\cite{N} we constructed normal and hybrid meson states  and studied kinematical relativistic effects at the quark and gluon level.  In this work we will estimate the size of corrections to the $\pi_{1}$  decays originating from the meson exchange forces between mesons. Since  particle number is not conserved and the particle momenta are of the  same order as their masses, this problem should be treated in a relativistic  formalism. In Sec.~II we apply the Lippmann-Schwinger equation for the  two coupled $\pi\rho$ and $\pi b_{1}$ channels. The first is the channel in  which a narrow $\pi_1(1600)$ exotic has been observed ~\cite{Adams:1998ff}, and the latter is the 
predicted dominant decay mode of the $\pi_1$. Then in Sec.~III we describe  the computational procedure of solving the resulting integral equations  and we present a discussion of the numerical results.    

\section{Final state interactions in $\pi_1$ decay}

\begin{figure}[ht]
\centering
\includegraphics[width=2.0in]{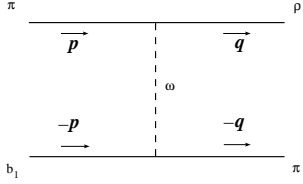}
\caption{\label{final} Final state interaction 
$\pi b_{1}\leftrightarrow\pi\rho$. }
\end{figure}
\begin{figure}[ht]
\centering
\includegraphics[width=0.9in,angle=270]{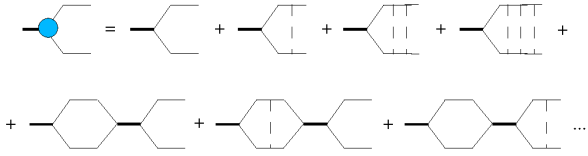}
\caption{\label{diag1} The lowest order FSI corrections to the $\pi_{1}$ 
decay from the interaction $\pi b_{1}\leftrightarrow\pi\rho$. A bold 
horizontal solid line represents a hybrid and normal horizontal solid lines 
refer to mesons. A vertical dashed line corresponds to a single $\omega$ 
meson exchange. The circle represents the overall decay amplitude. }
\end{figure} 

There exist several models of exotic meson decays, and all of them predict 
a small branching ratio to the $\pi\rho$ channel~\cite{IKP,CP,PSS}. It is 
possible, 
however, that this is modified by final state interactions between the 
outgoing mesons. The $b_{1}$ originating from the process 
$\pi_{1}\rightarrow\pi b_{1}$ can subsequently decay into $\pi$ and 
$\omega$, and the $\omega$ can absorb the other $\pi$ and produce a  
$\rho$, as shown in Fig.~\ref{final}. The $\omega$ exchange is expected 
to dominate the mixing potential since this is the dominant decay mode 
of the $b_1$ meson.    A similar mixing effect (mediated by $\rho$ 
exchange)  in connection with the $\eta\pi$ $P$-wave enhancement in the 
$1400\mbox{ MeV}$ mass region was considered by Donnachie and Page~\cite{Donnachie:1998ty}  in 
perturbation theory. However, since the meson couplings are large, one 
must sum up all possible amplitudes; with the lowest order diagrams shown 
in Fig.~\ref{diag1}.  In order to describe the total contribution of the 
final state interactions to the decay width of the $\pi_{1}$, we need to 
solve the Lippmann-Schwinger equation, 
\begin{equation}
T=V+VGT,
\end{equation}
which sums up the ladder of $\omega$ exchanges and the bare exotic meson 
exchanges.

We will denote the single particle exotic state $\pi_{1}$ by  
$|\alpha\rangle$, and the two-particle states $\pi\rho$ and $\pi b_{1}$ 
by Roman letters. 
We first introduce the matrix elements for the potential $V$,
\begin{equation}
V_{\alpha i}=\langle\alpha|V|i\rangle,\,\,V_{ij}=\langle i|V|j\rangle,
\end{equation}
and similarly for $T$. The elements $V_{\alpha i}$ are just the amplitudes 
of the corresponding "bare" decays of the $\pi_{1}$ describing transitions at 
the quark level, whereas $V_{ij}$ are related to the final state interaction 
potential.
In our state space the Lippmann-Schwinger equation can be written as,
\begin{eqnarray}
& & T_{\alpha i}=V_{\alpha i}+V_{\alpha j}G_{j}T_{ji}, \nonumber \\
& & T_{ji}=V_{ji}+V_{jk}G_{k}T_{ki}+V_{j\alpha}G_{\alpha}T_{\alpha i},
\label{LSE}
\end{eqnarray}
where 
\begin{equation}
G_{i}(E)=[E-H_{0}(i)+i\epsilon]^{-1}. 
\end{equation}

\begin{figure}[ht]
\centering
\includegraphics[width=1.0in,angle=270]{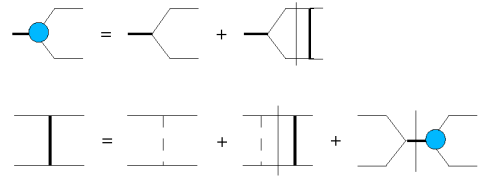}
\caption{\label{diag2} Diagrammatical representation of the 
Lippmann-Schwinger equation. A bold vertical solid line corresponds to 
the total amplitude of the interaction between two-meson states, whereas 
a normal vertical solid line represents the sum over all intermediate 
states. } 
\end{figure}  
\begin{figure}[ht]
\centering
\includegraphics[width=1.0in,angle=270]{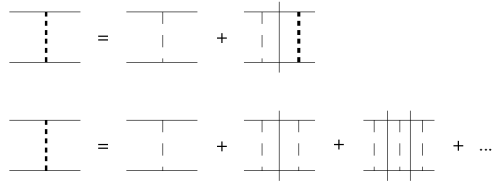}
\caption{\label{diag3} The Lippmann-Schwinger equation in a meson subspace. }
\end{figure}

Both equations in Eq.~(\ref{LSE}) are represented diagrammatically in 
Fig.~\ref{diag2}.
Hereinafter, if an index appears twice or more then a summation over this 
index is implicitly assumed. The Hamiltonian of the state $|i\rangle$ 
is $H_{0}(i)$, $\epsilon\rightarrow 0^{+}$, and $E$ is the energy which 
will be taken as the mass of the $\pi_{1}$. The matrix elements  $T_{ij}$ 
can be eliminated using the second equation in Eq.~($\ref{LSE}$), leading 
to  
 \begin{equation}
T_{\alpha i}=V_{\alpha i}+V_{\alpha j}G_{j}[1-VG]^{-1}_{jk}
(V_{ki}+V_{k\alpha}G_{\alpha}T_{\alpha i}).
\end{equation}
Now we introduce a subset of the $T$ matrix which acts only between 
states $|\pi\rho\rangle$ and $|\pi b_{1}\rangle$.  This is denoted by $t$ 
and is defined by
\begin{equation}
t=V+VGt,\,\,\,t_{ij}=[1-VG]^{-1}_{ik}V_{kj}.
\label{t}
\end{equation}
Diagrammatically this represents the sum of all diagrams without hybrid 
intermediate states, as shown in Fig.~\ref{diag3}.
Therefore we obtain 
\begin{eqnarray}
& & T_{\alpha i}=V_{\alpha i}+V_{\alpha j}G_{j}t_{ji}+ \nonumber \\
& & +V_{\alpha j}G_{j}[\delta_{jk}+(tG)_{jk}]
V_{k\alpha}G_{\alpha}T_{\alpha i}= \nonumber \\
& & =(V[1-GV]^{-1})_{\alpha i}+[V(1+Gt)GV]_{\alpha\alpha}
G_{\alpha}T_{\alpha i}.
\end{eqnarray}

\begin{figure}[t]
\centering
\includegraphics[width=0.5in,angle=270]{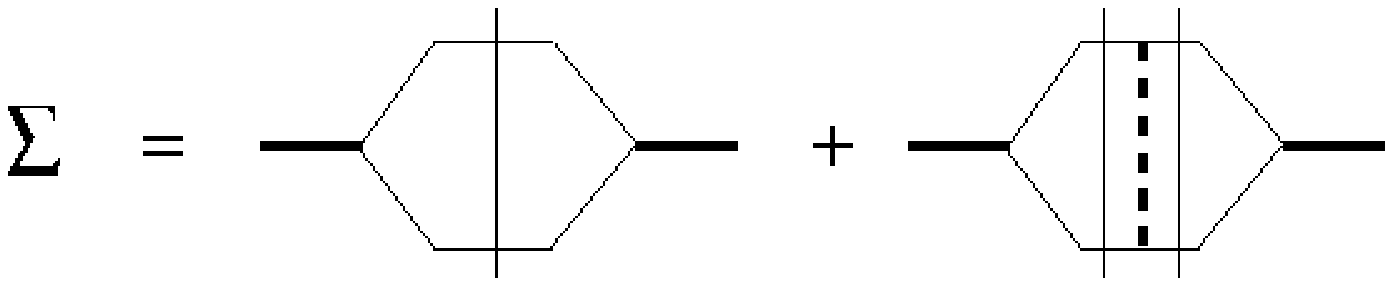}
\caption{\label{diag4} The hybrid self-energy. }
\end{figure}

\begin{figure}[t]
\centering
\includegraphics[width=0.5in,angle=270]{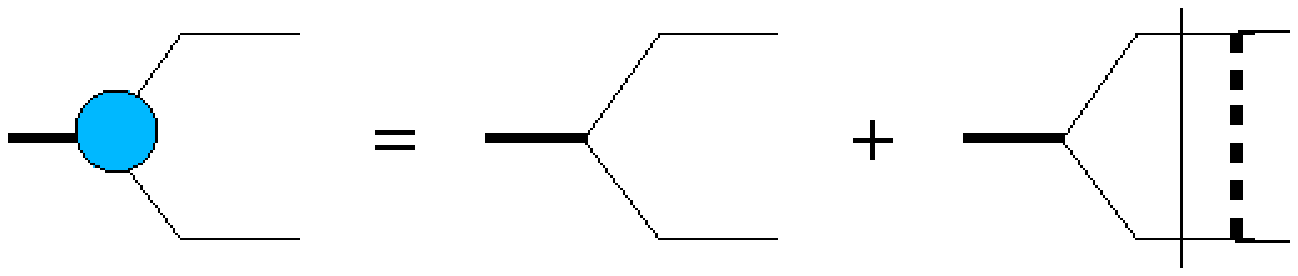}
\caption{\label{diag5} FSI correction to the $\pi_{1}$ decay amplitude. }
\end{figure}

Solving for $T_{\alpha i}$ gives,
\begin{equation}
T_{\alpha i}=G^{-1}_{\alpha}[G^{-1}_{\alpha}-\Sigma_{\alpha}]^{-1}
(V[1-GV]^{-1})_{\alpha i},
\end{equation}
where $\Sigma_{\alpha}=(V(1+Gt)GV)_{\alpha\alpha}$ is the self-energy of 
the $\pi_{1}$, shown in Fig.~\ref{diag4}.
Both $G^{-1}_{\alpha}$ and $\Sigma_{\alpha}$ are c-numbers, and thus
\begin{equation}
T_{\alpha i}=\frac{E-m_{ex}}{E-m_{ex}-\Sigma_{\alpha}}(V+VGt)_{\alpha i},
\end{equation}
where we again used $[1-GV]^{-1}=1+Gt$. A resonance production process has 
the form of $i \to \pi_1 X \to M_1 M_2 X$, where $i$ denotes an initial 
state and $X$ denotes possible other particles in the final state, assumed 
not to interact with the resonance $\pi_1$, and $M_1$ ($M_2$) is the final 
state meson (isobar) state. The full amplitude has the form 

\begin{eqnarray}
& & A(i \to \pi_1 X \to M_1 M_2 X)=P_\alpha G_\alpha  T_{\alpha i}= 
 \nonumber \\
& & =P_\alpha {1\over {E -m_{ex} - \Sigma_{\alpha}}} (V + V G t)_{\alpha i},
 \label{a}
 \end{eqnarray}
 where $P_\alpha$ describes the coupling between the bare exotic and the 
initial state. Thus in the full amplitude the bare exotic propagator 
$(E- m_{ex})^{-1}$ is replaced by the dressed one, 
$(E - m_{ex} - \Sigma_\alpha)^{-1}$. Furthermore if $\Sigma_\alpha$, which 
is energy dependent, is expanded around the physical exotic mass, one 
obtains the Breit-Wigner propagator. Here, we are interested in the 
effective coupling of the exotic to the two-meson channels. This is 
determined by the last term in Eq.~(\ref{a}). In the absence of FSI, the 
coupling is determined by the matrix $V$ which changes to $V + V G t$ due 
to re-scattering. We will therefore be studying the reduced $T$ matrix 
element, 

\begin{equation}
T_{\alpha i}=V_{\alpha i}+V_{\alpha j}G_{j}t_{ji},
\label{T}
\end{equation}
with $t_{ji}$ defined by ($\ref{t}$).
The diagrammatic representation of Eq.~(\ref{T}) is given in Fig.~\ref{diag5}.


Eq.~($\ref{t}$) written in a full notation in the rest frame of the 
$\pi_{1}$ is given by 
\begin{widetext}
\begin{equation}
t({\bf p},{\bf q},\lambda,\lambda')=
V({\bf p},{\bf q},\lambda,\lambda')+\sum_{\lambda"}\int\frac{d^{3}{\bf k}}
{(2\pi)^{3}4\omega_{1}(k)\omega_{2}(k)}V({\bf p},{\bf k},\lambda,\lambda")
G(k)t({\bf k},{\bf q},\lambda",\lambda'),
\label{t2}
\end{equation}
\end{widetext}
where 
\begin{eqnarray}
& & \omega_{i}(k)=E(m_{i},{\bf k}),\,\,\,\,H_{0}(k)=\omega_{1}(k)+
\omega_{2}(k), \nonumber \\
& & G(k)=[E-H_{0}(k)+i\epsilon]^{-1},
\end{eqnarray}
and
$m_{i},\,\,i=1,2$ are the masses of mesons in the intermediate two-meson 
state (related to $G$). In the above ${\bf p}$ and ${\bf q}$ are the 
relativistic relative three-momenta between two mesons, and the dependence 
on the center-of-mass momentum has been already factored out. Spins 
$\lambda$ refer to either $b_{1}$ or $\rho$.

We introduce the partial wave potentials by 
\begin{widetext}
\begin{equation}
V_{LL'}(p,q)=\sum_{M,M',\lambda,\lambda',j}\int d\Omega_{{\bf p}}
d\Omega_{{\bf q}}\langle L,M;1,\lambda|J,j\rangle \langle L',M';1,
\lambda'|J,j\rangle V({\bf p},{\bf q},\lambda,\lambda')Y_{LM}({\bf p})
Y^{\ast}_{L'M'}({\bf q}),
\label{part}
\end{equation}
\end{widetext}
where $d\Omega_{{\bf k}}$ is the element of the solid angle in the direction 
of the vector ${\bf k}$ and $k=|{\bf k}|$.
Similarly we define $t_{LL'}(p,q)$.
Substituting  $V_{LL'}(p,q)$ into ($\ref{t2}$) gives, 
\begin{widetext}
\begin{equation}
t_{LL'}(p,q)=V_{LL'}(p,q)+\sum_{L"}\int\frac{k^{2}dk}{(2\pi)^{3}
4\omega_{L",1}(k)\omega_{L",2}(k)}V_{LL"}(p,k)G_{L"}(k)t_{L"L'}(k,q),
\label{t3}
\end{equation}
\end{widetext}
where 
\begin{eqnarray}
& & \omega_{L",i}(k)=E(m_{L",i},{\bf k}), \nonumber \\
& & G_{L"}(k)=[E-\omega_{L",1}(k)-\omega_{L",2}(k)+i\epsilon]^{-1}.
\end{eqnarray}
In our state space we can have $L=0,2$ (the relative angular momentum 
between $\pi$ and $b_{1}$) or $L=1$ (between $\pi$ and $\rho$). For a 
$\pi_{1}$ we also have $J=1$. Thus in Eq.~(\ref{t3}) we must substitute:
\begin{equation}
m_{L,1}=m_{\pi},\,\,\,\,m_{0,2}=m_{2,2}=m_{b_{1}},\,\,\,\,m_{1,2}=m_{\rho}.
\end{equation}
Parity conservation reduced the number of non-vanishing matrix elements of 
the final state interaction potential to $V_{01}$, $V_{10}$, $V_{12}$ and 
$V_{21}$. Moreover, from CP invariance we have $V_{01}=V^{\ast}_{10}$ 
and $V_{12}=V^{\ast}_{21}$. The integral equation ($\ref{t3}$) cannot be 
solved analytically  and one needs to replace it by a set of matrix 
equations. The details will be given 
in the next section. When this is done and $t_{LL'}$ are found, we may go 
back to Eq.~($\ref{T}$) which becomes
\begin{eqnarray}
& & {\tilde{a}}_{L}(P)=a_{L}(P)+\sum_{L'}\int\frac{k^{2}dk}{(2\pi)^{3}
(2J+1)4\omega_{1}(k)\omega_{2}(k)} \nonumber \\
& & \times a_{L'}(k)G(k)_{L'}t_{L'L}(k,P),
\label{T2}
\end{eqnarray}
where $a_{L}(P)$ are the partial decay amplitudes defined in Ref.~\cite{N}. 
In order to obtain the corrected widths, they must be replaced by the 
corrected amplitudes ${\tilde{a}}_{L}(P)$.

Finally, we proceed to the form of the final state interaction potential.
The dominant effective interactions between the $b_1 \pi$ and the $\rho\pi$ 
channels are expected to originate from $\omega$ exchange. 
The $b_1 \to \omega \pi$ is the dominant decay channel of the $b_1$ meson 
and the $\rho\omega\pi$ coupling is also known to be large. 
The effective Lagrangian for the $\rho\pi\omega$ vertex is
\begin{equation}
L_{\rho\pi\omega}=g_{\rho\pi\omega}\epsilon^{\mu\nu\lambda\sigma}
\partial_{\mu}\omega_{\nu}\pi^{i}\partial_{\lambda}\rho^{i}_{\sigma},
\label{fsiL1}
\end{equation}
and for the $b_{1}\pi\omega$ vertex is
\begin{equation}
L_{b_{1}\pi\omega}=g_{b_{1}\pi\omega}\partial_{\mu}b_{1}^{\mu i}\pi^{i}
\omega_{\mu}. 
\label{fsiL2}
\end{equation}
Here the index $i$ corresponds to isospin.  The coupling constant, 
$g_{\rho\pi\omega}$ lies between 0.01 MeV$^{-1}$ and 0.02 MeV$^{-1}$ and we 
will take a value 0.014 MeV$^{-1}$ \cite{KL}, whereas 
 $g_{b_{1}\pi\omega}$, can be obtained by calculating the width of the 
decay $b_{1}\rightarrow\pi\omega$ and comparing with its experimental value 
of 142 MeV \cite{PDG}. Consequently, the amplitude for  
$b_{1}\rightarrow\pi\omega$ is equal to
\begin{equation}
A=-g_{b_{1}\pi\omega}\epsilon^{i}(\lambda_{b_{1}})\epsilon^{i\ast}
(\lambda_{\omega},{\bf P}),
\end{equation}
where ${\bf P}$ is given by
\begin{equation}
E(m_{\pi},{\bf P})+E(m_{\omega},{\bf P})=m_{b_{1}},
\end{equation}
and $\epsilon^i(\lambda)$ are the spin-1 polarization four-vectors.  
The absolute value of the square of this amplitude, summed over 
$\lambda_{\omega}$ and averaged with respect to $\lambda_{b_{1}}$, gives 
\begin{equation}
|A|^{2}=g^{2}_{b_{1}\pi\omega}(1+\frac{P^{2}}{3m_{\omega}^{2}}),
\end{equation}
and using  $\Gamma_{s-wave}=142$\,MeV$\cdot\,0.92=131$ MeV (the factor 
0.92 comes from the partial wave D/S ratio) we get $g_{b_{1}\pi\omega}=3650$ 
MeV. Therefore $g_{FSI}=g_{\rho\pi\omega}\cdot g_{b_{1}\pi\omega}$ is 
approximately  50.

The final state interaction potential can be obtained from Lagrangians 
of Eqs.~($\ref{fsiL1}$) and ($\ref{fsiL2}$) expressed in momentum space 
and dressed with the instantaneous $\omega$ propagator,
\begin{eqnarray}
& & V({\bf p},{\bf q},\lambda_{b_{1}},\lambda_{\rho})=g_{FSI}
\epsilon_{\mu\nu\sigma\tau}p^{\mu}q^{\nu}\frac{1}{({\bf p}-{\bf q})^{2}
+m^{2}_{\omega}} \nonumber \\
& & \times\epsilon^{\sigma}(\lambda_{b_{1}},-{\bf p})\epsilon^{\tau\ast}
(\lambda_{\rho},{\bf q}),
\label{FSIpot}
\end{eqnarray}
where ${\bf p}$ is the momentum of the $\pi$ in the $|\pi b_{1}\rangle$ 
state and ${\bf q}$ is the momentum of the $\rho$. At short distances, (for 
large values of $p$ and $q$) this potential is singular which is a 
consequence of treating mesons as elementary particles. Therefore we must 
regulate this potential with an extra factor that tends to zero for large 
momenta. We will choose an exponential function
\begin{equation}
e^{-|{\bf p}_{\omega}|/\Lambda},
\end{equation}
  where $\Lambda$ is a scale parameter expected to be on the 
order of the inverse of the meson radius $\Lambda \sim 0.5 - 1 \mbox{ GeV}$. 

\section{Numerical results}

The Lippmann-Schwinger integral equation of Eq.~(\ref{t3}) corresponds to 
outgoing wave boundary conditions. This means that the singularity of the 
term $G(k)$ is handled by giving the energy $E$ a small positive imaginary 
part $i\epsilon$. An integral of this form may be solved using the Cauchy 
principal-value prescription,
\begin{widetext}
\begin{eqnarray}
& & \int_{0}^{\infty}\frac{f(k)dk}{E-H_{0}(k)+i\epsilon}=
\wp\int_{0}^{\infty}\frac{f(k)dk}{E-H_{0}(k)}\,-\,i\pi f(k_{0})\biggl
(\frac{\partial H_{0}(k)}{\partial k}\biggr)^{-1}_{k=k_{0}}= \nonumber \\
& &  =\int_{0}^{\infty}\frac{dk}{k^{2}-k^{2}_{0}}\biggl[
\frac{f(k)(k^{2}-k^{2}_{0})}{E-H_{0}(k)}-\frac{2f(k_{0})\omega_{1}(k_{0})
\omega_{2}(k_{0})}{E}\biggr]-\,i\pi f(k_{0})\frac{\omega_{1}(k_{0})
\omega_{2}(k_{0})}{k_{0}E}.
\end{eqnarray}
\end{widetext}
Making a transition
\begin{equation}
f(k)\rightarrow \frac{k^{2}V(p,k)t(k,q)}{4(2\pi)^{3}\omega_{1}(k)
\omega_{2}(k)},
\end{equation}
and including a summation over partial waves, leads to a desired equation 
for $t_{LL'}(p,q)$ that no longer has a singularity:
\begin{widetext}
\begin{eqnarray}
& & t_{LL'}(p,q)=V_{LL'}(p,q)+\sum_{L"}\int_{0}^{\infty}
\frac{dk}{4(2\pi)^{3}(k^{2}-k^{2}_{0,L"})}\biggl[
\frac{k^{2}(k^{2}-k^{2}_{0,L"})V_{LL"}(p,k)t_{L"L'}(k,q)}{[E-H_{0,L"}(k)]
\omega_{L",1}(k)\omega_{L",2}(k)} \nonumber \\
& & +\frac{2k^{2}_{0,L"}}{E}V_{LL"}(p,k_{0,L"})t_{L"L'}(k_{0,L"},q)\biggr]
\,-i\pi\sum_{L"}\frac{k_{0,L"}V_{LL"}(p,k_{0,L"})t_{L"L'}(k_{0,L"},q)}
{4(2\pi)^{3}E},
\label{t4}
\end{eqnarray}
\end{widetext}
where 
\begin{equation}
H_{0,L}(k)=\omega_{L,1}(k)+\omega_{L,2}(k)
\end{equation}
and the quantities $k_{0,L}$ are defined by 
\begin{equation}
H_{0,L}(k_{0,L})=E.
\label{k0}
\end{equation}

In the decay $\pi_{1}\rightarrow\pi b_{1}$, the D-wave amplitude is 
negligible compared to the S-wave, thus we may reduce our angular 
momentum space to $L=0,1$. Therefore, we only need the formula for 
$V_{01}(p,q)$, and after expressing Clebsch-Gordan coefficients in 
Eq.~(\ref{part}) in terms of the spin-1 polarization vectors we obtain
\begin{eqnarray}
& & V_{01}(p,q)=-\frac{i\sqrt{3}}{4\pi}\sum_{\lambda,\lambda'}\int 
d\Omega_{{\bf p}}d\Omega_{{\bf q}}V({\bf p},{\bf q},\lambda,\lambda') 
\nonumber \\
& & \times\epsilon^{ijk}\epsilon^{i\ast}(\lambda)\epsilon^{j}(\lambda')q^{k}/q,
\label{v01}
\end{eqnarray}
where $V({\bf p},{\bf q},\lambda,\lambda')$ was given in Eq.~(\ref{FSIpot}).
The S-matrix is obtained from the $t$ matrix via
\begin{equation}
S_{LL'}=\delta_{LL'}-i\frac{\sqrt{k_{0,L}k_{0,L'}}}{16\pi^{2}E}t_{LL'}
(k_{0,L},k_{0,L'}).
\label{scat0}
\end{equation}
As a $2\times2$ unitary matrix it can be parametrized by two real scattering 
phase shifts $\delta_{0}$ and $\delta_{1}$, and one inelasticity parameter, 
$\eta$, ($0\leq\eta\leq1$), 
\begin{equation}
S_{LL'}=\left( \begin{array}{cc}
\eta e^{2i\delta_{0}} & i\sqrt{1-\eta^{2}}e^{i(\delta_{0}+\delta_{1})} \\
i\sqrt{1-\eta^{2}}e^{i(\delta_{0}+\delta_{1})} & \eta e^{2i\delta_{1}} 
\end{array} \right).
\label{scat1}
\end{equation}

Eq.~(\ref{t4}) may be solved, as we already mentioned, by converting the 
integration over $k$ into a sum over $N$ integration points 
$k_{n},\,\,n=1,2,...\,N$ (determined by Gaussian quadrature) with weights 
$w_{n}$ \cite{num}.



\begin{figure}[ht]
\centering
\includegraphics[width=1.5in,angle=270]{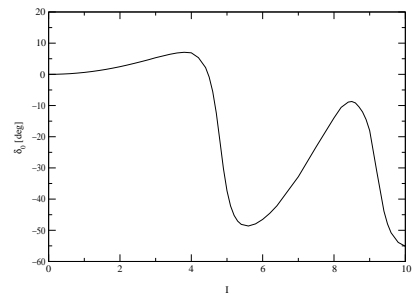}
\caption{\label{d0_v} Phase shift $\delta_{0}$ in degrees for the 
interaction $\pi b_{1}-\pi\rho$ as a function of the potential strength $I$, 
for $E=1.6$ GeV and $\Lambda=2$ GeV. }
\end{figure}
\begin{figure}[ht]
\centering
\vspace{0.27in}
\includegraphics[width=1.5in,angle=270]{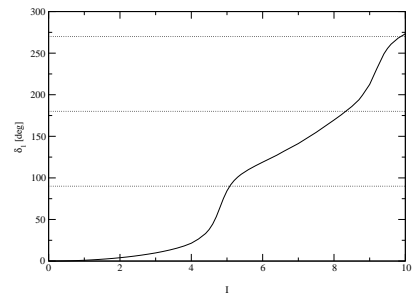}
\caption{\label{d1_v} Phase shift $\delta_{1}$ in degrees for the 
interaction $\pi b_{1}-\pi\rho$ as a function of the potential strength 
$I$, for $E=1.6$ GeV and $\Lambda=2$ GeV. }
\end{figure}
\begin{figure}[ht]
\centering
\includegraphics[width=1.5in,angle=270]{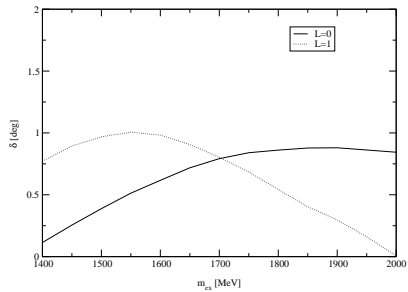}
\caption{\label{d01_mex} Phase shifts $\delta_{0}$ and $\delta_{1}$ in 
degrees, for the interaction $\pi b_{1}-\pi\rho$ as functions of the 
energy $E$, for $I=1$ and $\Lambda=2$ GeV. }
\end{figure}

Let us assume that the coupling constant $g_{FSI}$ is a variable, and we 
introduce the potential strength $I=g_{FSI}/g^{(0)}_{FSI}$, where 
$g^{(0)}_{FSI}=50$. In Figs.~\ref{d0_v} and \ref{d1_v} we present the 
phase shifts as functions of $I$, for the energy $E=1.6$ GeV and the scale 
parameter $\Lambda=2$ GeV. Their dependence on $E$ for $I=1$ and $\Lambda=2$ 
GeV is shown in Fig.~\ref{d01_mex}.


Within an experimental range of the expected hybrid mass $m_{ex}=E$ and the 
hadrodynamical coupling constant $g$, the phase shifts for both channels 
$L=0$ ($\pi b_{1}$) and $L=1$ ($\pi\rho$) are rather small. 
It is probably because $m_{ex}$ is large, thus the Born approximation is 
accurate. Furthermore, smaller values of $m_{ex}$ are closer to the 
$\pi b_{1}$ threshold, which (together with the second power of the FSI 
potential in momenta) should make the phase shifts small.  
Therefore, one expects that the final state interaction should not 
dramatically change the widths of $\pi_{1}\rightarrow\pi b_{1}$ and 
$\pi_{1}\rightarrow\pi\rho$ obtained without final state interactions, 
such as was discussed in Ref.~\cite{N}.

\begin{figure}[ht]
\centering
\includegraphics[width=1.5in,angle=270]{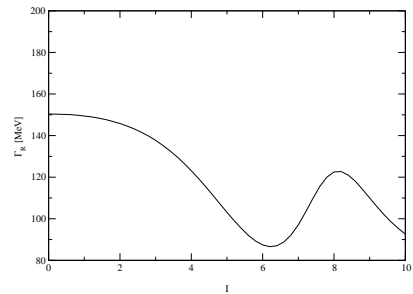}
\caption{\label{b1_fsi_v} FSI-corrected width of 
$\pi_{1}\rightarrow\pi b_{1}$ in MeV as a function of the potential 
strength $I$, for $m_{ex}=1.6$ GeV and $\Lambda=2$ GeV. }
\end{figure}

\begin{figure}[ht]
\centering
\includegraphics[width=1.5in,angle=270]{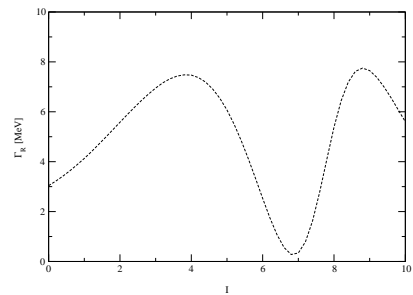}
\caption{\label{rho_fsi_v} FSI-corrected width of 
$\pi_{1}\rightarrow\pi\rho$ in MeV as a function of the potential 
strength $I$, for $m_{ex}=1.6$ GeV and $\Lambda=2$ GeV. }
\end{figure}

In Tables~\ref{fsi1} and~\ref{fsi2} we compare the original widths (in MeV) 
with the FSI-corrected ones for various values of $m_{ex}$ and $\Lambda$, 
for $I=2$. The strength was doubled in order to compensate for the effect 
of the regulating factor in the FSI potential. For the values of $\Lambda$ 
other than 2 GeV, the coupling constant $g_{FSI}$ was renormalized such 
that the value of $V_{01}(p_{0},q_{0})$ remained constant. Here $p_{0}$ 
and $q_{0}$ are the relative momenta between mesons for the $\pi b_{1}$ 
and $\pi\rho$ channels, respectively. In Figs.~\ref{b1_fsi_v} and 
\ref{rho_fsi_v} we show their dependence on $I$ for $m_{ex}=1.6$ 
GeV and $\Lambda=2$ GeV. The behavior of $\delta_{1}$ indicates the 
existence of $\pi\rho$ resonances for the free parameters given. 

From Figs.~\ref{b1_fsi_v} and~\ref{rho_fsi_v} it follows that the 
$\pi b_{1}$ channel can change significantly, if we increase the potential 
strength enough. However, even for large potentials the $\pi\rho$ channel 
width remains on the order of a few MeV. These results, which are shown 
in Tables~\ref{fsi1} and~\ref{fsi2}, are only weakly dependent on the 
value of the parameter $\Lambda$. Therefore, final state interactions 
cannot produce a large width for the $\pi_1 \rightarrow \pi\rho$ mode.
 
\begin{table}[ht]
\centering
\begin{tabular}{|c||r|r|r|r|}
\hline
$m_{ex}$\,[GeV] & no FSI & $\Lambda=0.5$ & $\Lambda=1$ & $\Lambda=2$ [GeV]\\
\hline\hline
1.4 & 85 & 84 & 86 & 87\\
\hline
1.5 & 153 & 149 & 151 & 153\\
\hline
1.6 & 150 & 144 & 145 & 145\\
\hline
1.7 & 124 & 120 & 118 & 117\\
\hline
1.8 & 95 & 92 & 89 & 87\\
\hline
1.9 & 69 & 67 & 65 & 62\\
\hline
2.0 & 48 & 47 & 45 & 43\\
\hline
\end{tabular}
\caption{\label{fsi1} Original and FSI-corrected widths in MeV for the 
$\pi b_{1}$ mode of the $\pi_{1}$ decay, for various values of $m_{ex}$ 
and $\Lambda$. }
\end{table}

\begin{table}[ht]
\centering
\begin{tabular}{|c||r|r|r|r|}
\hline
$m_{ex}$\,[GeV] & no FSI & $\Lambda=0.5$ & $\Lambda=1$ & $\Lambda=2$ [GeV]\\
\hline\hline
1.4 & 3 & 8 & 8 & 8\\
\hline
1.5 & 3 & 6 & 7 & 7\\
\hline
1.6 & 3 & 4 & 5 & 6\\
\hline
1.7 & 2 & 2 & 3 & 4\\
\hline
1.8 & 2 & 1 & 2 & 3\\
\hline
1.9 & 2 & 1 & 1 & 2\\
\hline
2.0 & 1 & 1 & 1 & 1\\
\hline
\end{tabular}
\caption{\label{fsi2} Original and FSI-corrected widths in MeV for the 
$\pi\rho$ mode of the $\pi_{1}$ decay, for various values of $m_{ex}$ 
and $\Lambda$. }
\end{table}

In Table~\ref{fsi3} we present the widths for both modes, assuming that 
the FSI potential was replaced by a simple separable potential,
\begin{equation}
V(p,q)=-V_{0}e^{-(p^{2}+q^{2})/\Lambda^{2}},
\label{toy}
\end{equation}
where $\Lambda$ is a free parameter on the order of 1 GeV. The strength 
$V_{0}$ was chosen such that the squares of the potentials in 
Eqs.~(\ref{v01}) and (\ref{toy}) were equal, when integrated over $p$ and 
$q$. We see that the results obtained using this toy potential with a 
comparable strength are quite similar to those for our full model. 

\begin{table}[ht]
\centering
\begin{tabular}{|c||r|r|}
\hline
$m_{ex}$\,[GeV] & $\pi b_{1}$ & $\pi\rho$\\
\hline\hline
1.4 & 83(85) & 9(9)\\
\hline
1.5 & 145(149) & 7(7)\\
\hline
1.6 & 142(144) & 5(5)\\
\hline
1.7 & 118(117) & 3(4)\\
\hline
1.8 & 91(89) & 2(2)\\
\hline
1.9 & 66(64) & 1(1)\\
\hline
2.0 & 46(44) & 1(1)\\
\hline
\end{tabular}
\caption{\label{fsi3} FSI-corrected widths in MeV for the $\pi b_{1}$ and 
$\pi\rho$ modes of the $\pi_{1}$ decay, for various values of $m_{ex}$ 
and for $\Lambda=$\,1 GeV (2 GeV). }
\end{table}

\section{Summary}

We have found that only unnaturally large potentials in the final state may 
change the widths for exotic meson decays calculated in microscopic models. 
In our model, the partial widths tend to change by only a few MeV. 
However, the $\pi\rho$ channel is subjected to a rather large relative 
correction. It may be caused by a large value of the original width for 
the $\pi b_{1}$ mode. The net width for the $\pi_{1}\rightarrow\pi\rho$ 
channel is still of order of few MeV which agrees with the results of Refs.\ 
\cite{CP} or \cite{PSS}, but disagrees with results of Ref.\  
\cite{CD}.


\begin{acknowledgments}
 This work was supported in part by the US Department of 
Energy under contract DE-FG0287ER40365 and National Science Foundation 
grant NSF-PHY0302248.
\end{acknowledgments}

\end{document}